\documentclass[prl,twocolumn,showpacs,floatfix,aps]{revtex4}
\usepackage{graphicx}
\usepackage{bm}% bold math
\begin{document}
\preprint{cond-mat/0000000}
\title{Energy gap and proximity effect in $MgB_2$ superconducting wires}
\author{R.~Prozorov}
%\email{prozorov@mailaps.org}
\altaffiliation[Present address: ]{Department of Physics and Astronomy, University of South Carolina, SC
29208} \email{prozorov@mailaps.org}
\author{R.~W.~Giannetta}
\affiliation{Loomis Laboratory of Physics, University of Illinois at Urbana-Champaign, 1110 West Green St.,
Urbana, Illinois 61801.}
\author{S.~L.~Bud'ko}
\author{P.~C.~Canfield}
\affiliation{Ames Laboratory and Department of Physics and Astronomy, Iowa State University, Ames, IA 50011.}
\date{\today}

\begin{abstract}
Measurements of the penetration depth $\lambda (T,H)$ in the presence of a DC magnetic field were
performed in $MgB_2$ wires. In as-prepared wires $\lambda (T,H<130~Oe)$ shows a strong diamagnetic
downturn below $\approx~10~K$. A DC magnetic field of $130~ Oe$ completely suppressed the downturn. The
data are consistent with proximity coupling to a surface $Mg$ layer left during synthesis. A theory for
the proximity effect in the clean limit, together with an assumed distribution of the $Mg$ layer
thickness, qualitatively explains the field and temperature dependence of the data. Removal of the $Mg$
by chemical etching results in an exponential temperature dependence for $\lambda (T)$ with an energy
gap of $2 \Delta (0)/T_c\approx 1.54$ ($\Delta(0)~\approx~2.61~meV$), in close agreement with recent
measurements on commercial powders and single crystals. This minimum gap is only 44\% of the BCS weak
coupling value, implying substantial anisotropy.
\end{abstract}
\maketitle
Superconducting $MgB_2$ \cite{nagamatsu} presents, for possibly the first time, a combination of
phonon-mediated pairing together with a relatively high transition temperature ($T_c \approx 39.4~K$ )
comparable to hole-doped cuprates. Evidence for a phonon mechanism has come from several measurements
which indicate a substantial isotope effect \cite{canfield1, lawrie}. Tunnelling measurements have given
values of the energy gap ratio $\delta \equiv 2\Delta(0)/T_c$ ranging from $1.25$ to $4$
\cite{sharoni,karapetrov,schmidt,rubio}. NMR measurements of the $^{11}$B nuclear spin-lattice
relaxation rate give $\delta \approx 5$ \cite{kotegawa} while photoemission spectroscopy gives $\delta
\approx 3$ \cite{takahashi}. A recent tunnelling measurement has shown evidence for two energy gaps
\cite{szabo}.

The temperature dependence of the London penetration depth $\lambda$ is a sensitive probe of the
quasiparticle density of states and thus the minimum energy gap. Some early data for $\lambda$ in
commercial $MgB_2$ powders showed apparent power law behavior suggesting nodes
\cite{panagopoulos,pronin}. It is important to make sure that no extrinsic factors exist that may bias
the interpretation of $\lambda(T)$. A persistent complication has been the presence of surface
contaminants remaining from the growth process, most notably elemental $Mg$. In this paper we report
magnetic screening measurements of dense $MgB_2$ wires grown around a tungsten core. The presence of a
$Mg$ layer on as-grown wires gives rise to a large increase in the diamagnetic response below $10~K$.
The temperature and field dependencies of the magnetic screening are consistent with proximity induced
correlations. After etching, the same wires show exponential behavior with a gap ratio of $\delta
\approx 1.54$, less than 1/2 the BCS value of $3.53$ and very close to the value obtained from recent
penetration depth measurements on both commercial powders \cite{manzano,chen} and single crystals
\cite{carrington}.

Growth of the $MgB_2$ wires has been described in detail elsewhere \cite{canfield2}. In brief, boron
fibers and Mg with a nominal ratio of $MgB_2$ were sealed in a Ta tube. The tube was sealed in quartz
and placed in a box furnace at $950^oC$ for approximately two hours. The reaction ampoule was then
removed from the furnace and quenched to room temperature. The wire samples used here had a tungsten
core of $15~\mu m$ diameter, outer diameters of $180~\mu m$ and $200~\mu m$ and were $2~mm$ long. SQUID
magnetometer measurements showed essentially ideal Meissner screening ($ - 4\pi \chi = 1$) in applied
fields up to 1000 Oe. However, tunnel diode measurements with much higher sensitivity revealed a clear
diamagnetic downturn below $10~K$ which we show was due to surface $Mg$. The $Mg$ layer was identified
by local XRD analysis and could be etched away with an 0.5 \% solution of $HCl$ in ethanol. SEM pictures
after etching revealed a sinter of hexagonal $MgB_2$ crystallites with some traces of $MgO$.

The penetration depth was measured with an 11 MHz tunnel-diode driven LC resonator used in several
previous studies \cite{prozorov}. An external $dc$ magnetic field ($0-7~kOe$) could be applied parallel
to the $ac$ field ($\sim 5$ mOe) using a compensated superconducting solenoid. The oscillator frequency
shift $\Delta f = f (T)- f(T_{min})$ is proportional to the rf susceptibility and thus to changes in the
penetration depth, $\Delta \lambda = \lambda (T) - \lambda(T_{min})$ via $\Delta f = - G \Delta
\lambda$, where $G$ is a calibration constant \cite{prozorov}. The random orientation of $MgB_2$
crystallites implied that these measurements represent an average over in-plane and out-of-plane
$\lambda$. The polycrystalline nature also made it difficult to reliably estimate G for the wires.
Therefore, all penetration depth data is plotted as raw frequency shift, after subtraction of the sample
holder background. Decreasing frequency corresponds to increased diamagnetic screening.
\begin{figure} [htb]
\includegraphics[width=8.5cm,keepaspectratio=true]{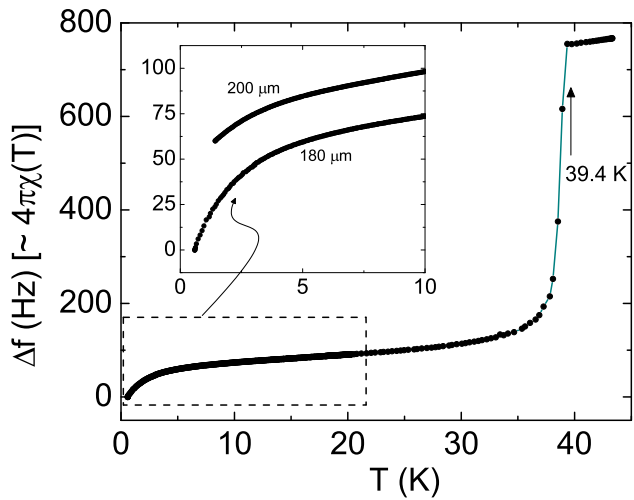}
\caption{Temperature dependence of the penetration depth at zero DC field. \textit{Inset:}
low-temperature region for two different diameter wires. \label{fig1}}
\end{figure}

Figure \ref{fig1} shows the temperature dependence of the oscillator frequency in as-prepared wires for
zero DC field. The inset is a magnification of the low-temperature behavior, showing a pronounced
diamagnetic downturn below $10~K$ for two separate wires of somewhat different diameter. This downturn
disappeared completely upon etching the wires. We attribute this downturn to a proximity effect induced
in the $Mg$ surface layer. Enhanced diamagnetism is a generic feature of proximity systems as carriers
in the normal metal layer gradually acquire pairing correlations and develop a Meissner screening
response \cite{claassen,anlage,mota,nagano}.

Several quasiclassical analyses of proximity systems ranging from clean to intermediate have shown that
the characteristic temperature for the appearance of screening is $5~T_A$ where $T_A  = \hbar V_F /2\pi
k_B d$ is the Andreev temperature \cite{zaikin,blatter,belzig}. For clean systems, $T_A$ is the
temperature at which the normal metal coherence length $\xi _N(T)=\hbar V_F/2\pi k_B T$ equals the
normal metal layer thickness $d$. Here $V_F$ is the Fermi velocity. For $Mg$ the coherence length varies
from $0.2~ \mu m$ at $T = 1~K$ to $2~ \mu m$ at $10~ K$, suggesting an average $Mg$ thickness of $2~\mu
m$ and an Andreev temperature $T_A \approx 0.6~K$.

Whether the proximity sandwich is in the clean or dirty limit depends upon the electronic mean free
path, $\ell_e$, in the $Mg$ layer, which is not known accurately. For example, a residual resistance
ratio of 20 would give $\ell_e = 0.2~\mu m$, which must be compared to both $\xi_N$ and $d$. The clean
limit requires $\ell_e  >> \min \left\{ {\xi _N ,d} \right\}$ while the dirty limit requires $\ell_e  <
< \xi_N ,d$. Strictly speaking, the latter regime requires that both the normal metal \textit{and}
superconductor be in dirty limit, which is most likely not true for $MgB_2$ \cite{belzig}. Over the
temperature range $1 - 10~ K$, all three numbers are comparable and we are likely in an intermediate
range for which there is no analytic solution for the susceptibility \cite{belzig}. Uncertainties in the
parameters did not justify fitting to the full numerical solutions. We therefore fit the data to both
clean and dirty limits where analytic solutions are available in order to gain some qualitative
understanding.

In the clean limit the diamagnetic susceptibility of the normal metal layer is given by \cite{blatter},

\begin{equation}
4\pi \chi_{clean}= - \frac{3}{4}\frac{1}{1 + 3 \lambda _N^2(T)/d^2} \label{chiclean}
\end{equation}

The factor 3/4 comes from the nonlocal response in the normal metal layer that overscreens the external
field and $\lambda _N(T)$ is a length scale given by \cite{zaikin,blatter,belzig},

\begin{equation}
\frac{{\lambda _N \left( 0 \right)}}{{\lambda _N \left( T \right)}} = \gamma \left( {\Delta ,T}
\right)\sqrt {\frac{{6\xi _N \left( T \right)}}{d}} e^{ - d/\xi _N \left( T \right)} \label{lambda}
\end{equation}
Here $\lambda _N \left( 0 \right) = \sqrt {4\pi ne^2 /m} \approx 180$ \AA~ is formally the London
penetration depth in a superconductor with the carrier mass and density of $Mg$. The energy gap of the
superconductor enters through the factor,

\begin{equation}
\gamma \left( {\Delta ,T} \right) = \Delta /\left[ {\pi k_B T + \sqrt {\Delta ^2  + \left( {\pi k_B T}
\right)^2 } } \right]
 \label{gamma}
\end{equation}

We take $\Delta=2.6~meV$ from our own data shown later. For the dirty limit, a solution of the Usadel
equations \cite{usadel,narikiyo} leads to a power law susceptibility without any characteristic
temperature.

\begin{equation}
 - 4\pi \chi _{dirty}  \propto \frac{{\xi _D }}{d} = \frac{1}{6}\sqrt {\frac{{\hbar V_F \ell _e }}{{6\pi k_B T}}}
\label{chidirt}
\end{equation}

A key feature of the proximity effect is the disappearance of screening in applied fields greater than a
breakdown field $H_b (T,d)$. In the clean limit this field is given by \cite{blatter,degennes},

\begin{equation}
H_b(clean) \approx \frac{\sqrt 2 }{\pi }\gamma (\Delta, T)\frac{\phi _0}{\lambda _N \left( 0 \right) d}e^{ -
{d/\xi _N(T)}}
\label{breakdownclean}
\end{equation}

where $\phi_0$ is the superconducting flux quantum and the result holds for $T > > T_A $ The temperature
and film thickness dependence of $H_b (T,d)$ has been verified in proximity systems that vary from
somewhat dirty to clean \cite{mota} and should therefore be applicable here. The breakdown field in the
dirty limit is given by \cite{blatter,degennes},

\begin{equation}
H_b (dirty) \approx 1.9\frac{{\phi _0 \ell _e }}{{\lambda _N (0) \xi _D^2 }}e^{ - d/\xi _D }
\label{breakdowndirty}
\end{equation}
\begin{figure} [htb]
\includegraphics[width=8.5cm,keepaspectratio=true]{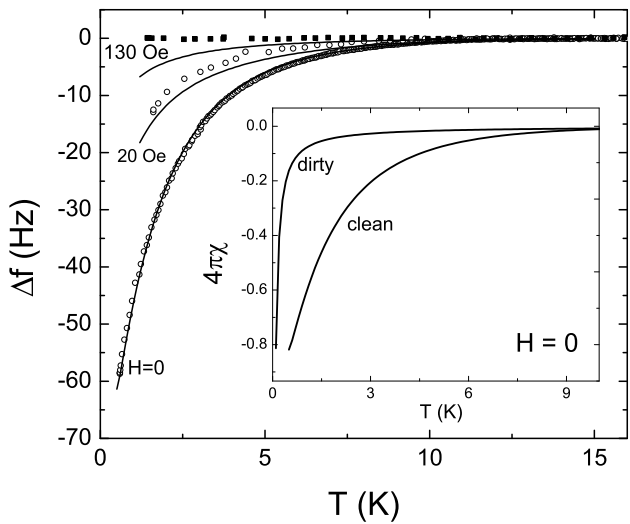}
 \caption{ Fits to clean limit proximity effect. Data in zero field were fit to
 obtain $d \approx 1.95~ \mu m$ and $\sigma \approx 1.2~ \mu m$. Finite field (solid) curves were
 then generated from Eq.~\ref{chi}. Data for $H = 0, 20$ and $130~Oe$ are shown. {\it Inset}:
 Comparison of dirty and clean limits at zero field. Dirty limit did not fit the data at any choice
 of parameters.\label{fig2}}
\end{figure}
Our device measures the screening of a very small AC field in the presence of a much larger DC field H.
We assume that once $H > H_b $ the AC screening vanishes. For $H < H_b $ we assume that the zero-field
susceptibility expression holds. This approach clearly ignores nonlinear effects which a more carefully
controlled experiment could address. The $Mg$ layer was not uniform and we used a probability
distribution for the normal metal layer thickness. The frequency shift measured upon extraction of the
sample from the coil {\it in-situ}, combined with the additional diamagnetic screening, Fig.~\ref{fig2},
gives an estimate of the $Mg$ layer thickness of $d \approx 1.62$, close to the clean limit fit value,
$d \approx 1.95$. Based upon many studies of film growth and random processes in condensed matter
systems we adopt a log-normal distribution of film thicknesses $p(x,d,\sigma)=(\sqrt{2 \pi} x
\sigma)^{-1} \exp{[(\log{x}-d)/\sqrt{2}\sigma]^{-2}}$ where $d$ is the mean thickness and $\sigma$ the
variance. A Gaussian distribution gave a less satisfactory fit. The proximity-enhanced diamagnetic
contribution to the signal was then taken to be,

\begin{equation}
\Delta f \propto \int_0^\infty  {\chi _N \left( T \right)p\left( {x:d,\sigma } \right)} \theta \left( {H -
H_b \left( {T,x} \right)} \right)dx
 \label{chi}
\end{equation}

In order to fit the data, we subtracted the signal from the superconductor alone, obtained after
etching. (The superconducting signal has negligible temperature dependence below $5~K$.) Data for $H =
0$ were then fit to Eq.~\ref{chi} with an overall scale factor, average thickness $d$ and $\sigma$ as
fitting parameters. (For the dirty limit we must also assume a mean free path.) We obtained $d = 2~\mu
m$ and $\sigma = 1.2~\mu m$. These parameters were then held fixed and the response in a finite magnetic
field was calculated. Fig.~\ref{fig2} shows the data and generated curves for $H = 0,~20,$ and $130~Oe$.
The finite field curves generated from the clean limit model all showed somewhat more screening than the
data. This is partly due to our assumption of a distribution of the $Mg$ layer thickness. Regions with
thickness smaller than the average will have higher breakdown fields and will continue to screen even
large DC fields. We were not able to find a satisfactory fit to the dirty limit. The inset shows the
best fit to the dirty limit that could be achieved. Given the uncertainty in parameters and the
crudeness of the model, we feel that the agreement with the clean limit model is reasonably good.
\begin{figure} [htb]
\includegraphics[width=8.5cm,keepaspectratio=true]{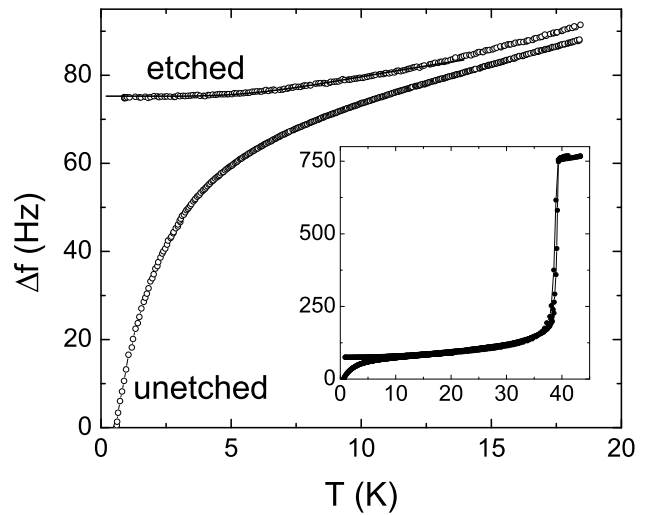}
 \caption{Main figure: $\Delta f(T)$ before and after etching. Thin solid line shows a BCS fit as
 described in  the text. Inset: full temperature scale $\Delta f(T)$ for etched and unetched wire.
 \label{fig3}}
\end{figure}
Both clean and dirty limits predict that the proximity effect will exhibit substantial hysteretic
effects. We observed no hysteresis for the range of fields shown here. This may be due to broadening of
the first order transition by the spread of film thicknesses. Hysteresis was observed at much higher
fields, of order $1500~Oe$. In this field we expect any proximity effect to be quenched but vortices
will be present in $MgB_2$. The hysteresis is then most likely due to trapped flux. We defer a
discussion of the higher field data to another paper.

In an effort to determine the pairing symmetry of pure $MgB_2$ the $Mg$ layer was etched away. The
result is shown in Fig.~\ref{fig3}. The downturn disappeared completely and the temperature dependence
became exponential. The inset shows full scale transition curves for both and etched and un-etched
wires. The transition temperature remained unchanged and the only apparent change due to etching is the
disappearance of the low-temperature diamagnetic downturn in $\lambda$.

\begin{figure} [htb]
\includegraphics[width=8.5cm,keepaspectratio=true]{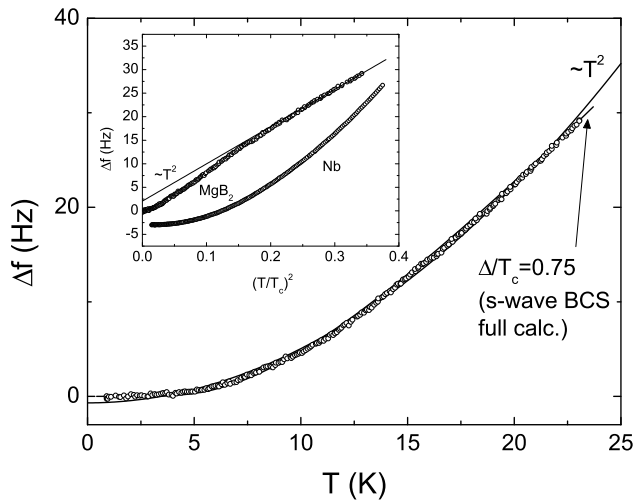}
 \caption{Weak-coupling s-wave BCS fit (full temperature range calculation) and quadratic power law fit.
 {\it Inset}: $MgB_2$ data vs $(T/T_c)^2$ along with polycrystalline Nb for comparison.\label{fig4}}

\end{figure}

Fig.~\ref{fig4} shows fits to both a full calculation (not low temperature expansion) of weak-coupling
s-wave BCS form for $\lambda(T)$ and to a quadratic power law. The BCS fit gives a value of $2
\Delta_0/T_c \approx 1.54$ ($2.6~meV$) which is 0.43 times the weak-coupling BCS ratio of 3.52, implying
a substantial anisotropy if the material is in the weak-coupling limit. The inset shows the data on a
$(T/T_c)^2$ scale along with data for polycrystalline $Nb$ foil, which gives an extremely good fit to
isotropic BSC theory. The quadratic power law gives a poorer fit and argues against a nodal order
parameter. Our value of $2.6~meV$ for the energy gap is in close agreement with recent penetration depth
measurements on commercial powders \cite{manzano} and single crystals \cite{carrington} which gave a
value of $2.8~meV$, and with tunnelling measurements which claimed a two band picture \cite{szabo}.

In conclusion, we have reported measurements of the magnetic penetration depth in dense $MgB_2$ wires.
We interpret the diamagnetic downturn in the effective $\lambda (T)$ for unetched wires as evidence for
a clean limit proximity effect between $MgB_2$ and an $Mg$ surface layer. After removing this $Mg$
layer, the results are consistent with a minimum gap value of $2.6~meV$.

\begin{acknowledgments} We thank A.~Carrington, J.~R.~Clem, D.~K.~Finnemore, R.~A.~Klemm and D.~Lawrie for
useful discussions. Work at UIUC was supported by the National Science Foundation, grant DMR-0101872.
Work at Ames was supported by the Director for Energy Research, Office of Basic Energy Sciences. Ames
Laboratory is operated for the U. S. Department of Energy by Iowa State University under Contract No.
W-7405-Eng.-82.
\end{acknowledgments}

\end{document}